# An entropic understanding of flow maldistribution in thermally isolated parallel channels

Toochukwu Aka, Shankar Narayan

Department of Mechanical, Aerospace, and Nuclear Engineering, Rensselaer Polytechnic Institute, 110 8th Street, Troy, NY 12180

*Corresponding Author: narays5@rpi.edu

ABSTRACT

Flow across heated parallel channel systems exists in many applications. The performance of such systems experiencing multiphase flow could suffer from the deleterious effects of flow non-uniformity or maldistribution. Modeling the behavior of such systems is challenging due to the inherent non-linearity associated with the multiphase flow and the difficulty in determining the actual flow among several possible flow distributions. This study addresses the challenge by analyzing the entropy production in such systems. Using experiments on two thermally isolated, nominally identical, and externally heated parallel channels, we quantify irreversibility in the resulting multiphase flow by evaluating the entropy generation rate. Our experiments reveal that certain flow conditions result in severe maldistribution (flow ratio > 10) in the channels, associated with a sharp rise in entropy production. Such an increase is not predicted for uniform flow distribution across parallel channels, making maldistributed flow a thermodynamically favored state over equally distributed flow. We extend this understanding to non-identical parallel channels as well. With entropy analysis providing additional insight besides the fundamental equations governing mass, momentum, and energy conservation, this approach is valuable in predicting and controlling flow distribution in parallel channel systems.

INTRODUCTION

Parallel channels can scale performance in several engineering systems and devices, such as heat exchangers, chemical reactors, electronic cooling and power generation systems, fluidized beds, microfluidic devices, and fuel cells [1]–[6]. Under certain operating conditions, such systems face the risk of instability, wherein flow distribution between parallel channels becomes severely non-uniform. This phenomenon could occur even for parallel channels that are nominally identical. It is undesirable since it can reduce the system's performance and sometimes cause irreversible damage [1][2]. Therefore, prior efforts have focused on the prediction and characterization of flow maldistribution in parallel channels, especially where multiple phases (liquid and vapor) coexist [2], [3], [7]–[11]. Taitel et al [10] showed that symmetrical heating of a multi channel system can be unstable and established a control procedure against flow maldistribution. Minzer et al [12] used linear stability analysis to differentiate between stable and unstable flow solutions. Results from Minzer's study, show that flow distribution may be dependent on history. Yang et al [13] developed a 3D computational model of a nine parallel channel system and showed that under the same conditions, flow tend to be more equally distributed in parallel channels with a circular crossection when compared to a triangular crossection. Zhang et al [7] with the aid of a model, showed that flow distribution in a parallel channel can be controlled by manipulating the total flow rate if the parallel channels are distinct. Natan et al [14] showed that up to five solutions may be obtained for a single flow rate through a two parallel channel system and speculated that the most feasible solution is the one with the least pressure drop.

In these studies, the methodology employed in predicting the stability of different flow distributions involves the perturbation of the Navier Stokes' equations[7], [10], [12], which establishes a criterion to ascertain if maldistribution can occur or not. While this approach distinguishes between stable and unstable flows [9], [12] it is unclear why a particular flow distribution is bound to occur over other mechanically feasible possibilities. We address this gap in knowledge by introducing thermodynamic considerations to flow distribution in parallel channels.

Entropy analysis, based on the second law of thermodynamics, is a valuable tool for analyzing the directionality of processes in physical systems. Entropy analysis evaluates and optimizes thermal-hydraulic systems by quantifying the degree of irreversibility associated with flow and energy transfer in systems. A recent study applied multiscale entropy analysis in identifying flow regimes and dynamic characteristics of gas-liquid two-phase flows in horizontal channels [15]. Other studies have focused on optimizing heat exchangers via entropy generation minimization [11] – [14]. This study uses entropy analysis to show the relationship between entropy generation and flow distribution in a system consisting of two thermally isolated parallel channels. In this case, externally heated channels cause vaporization and multiphase (liquid-vapor) flow. This study provides, for the first time, a thermodynamic perspective on flow distribution in parallel channels in such conditions. We demonstrate how the entropy analysis can be applied to distinguish between stable and unstable flow states corresponding to the same system operating conditions.

## FLOW DISTRIBUTION IN PARALLEL CHANNELS

The changes in flow properties like pressure and temperature across parallel channels are useful in quantifying the flow performance. For given inlet conditions, such as inlet pressure $P_i$ and temperature $T_i$, the exit properties like pressure $P_e$ and enthalpy $h_e$ can be calculated using the equations for conservation of mass, momentum balance, and the first law of thermodynamics. These fundamental equations take the following form for two thermally isolated parallel channels sharing a common inlet and exit (Figure 1) and operating at steady-state.

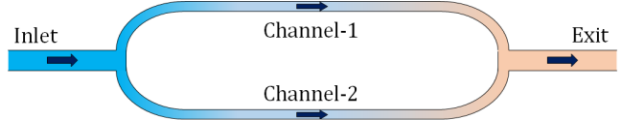

Figure 1. Two thermally isolated parallel channels sharing a common inlet and exit. The flow in the parallel channels experiences heating, which can result in boiling (phase-change) and vapor flow at the exit.

$$\int_{CS} \rho \bar{v} \cdot d\bar{A} = 0 \tag{1}$$

$$\int_{CS} \bar{v} \rho \bar{v} \cdot d\bar{A} = \int_{CS} -p \, d\bar{A} + \int_{CS} \bar{\bar{\tau}} \, d\bar{A} + \int_{CS} \bar{g} \, dV \tag{2}$$

Here $\bar{v}$, $p$, $\bar{\bar{\tau}}$, and $\bar{g}$ denote velocity, pressure, tangential stress, and body force per unit volume, respectively. In addition, the first law of thermodynamics gives

$$\int_{CS} h \rho \bar{v} \cdot d\bar{A} = \int_{CS} d\dot{Q} \tag{3}$$

Application of these fundamental equations to flow in individual channels allows predicting the pressure change, $\Delta P = P_i - P_e$ for different flow rates, $\dot{m}$. The model outcome could suggest multiple possible flow distributions for a simple case involving two nominally identical channels (Figure 1), where the flow is heated as it passes through them. For example, one possible model prediction can be equally distributed flow for all possible flow rates through the channels (Figure 2, bottom). Another possibility is that the flow is equally distributed only at high flow rates, whereas it is unequally distributed or maldistributed at low flow rates (Figure 2, top). A fundamental aspect still unclear is which of the two distributions is likely to occur and why.

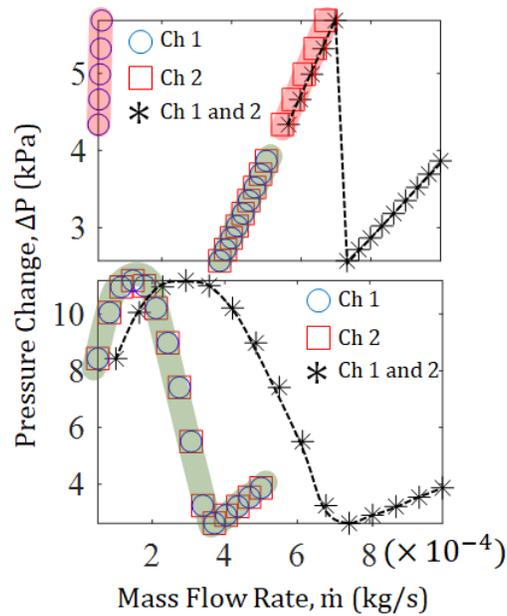

Figure 2. Possible flow distribution predictions in two parallel channels (fluid: water, diameter: $1.4 \times 10^{-3}$m, length: 0.3m ) satisfying the equations governing mass, momentum, and energy conservation. Circles (Ch 1) and squares (Ch 2) denote flow rates and pressure changes across individual channels. Stars with dotted lines denote flow rate and pressure change across both channels collectively. Models may either predict the equal distribution of flow at large flow rates (top, shaded green) and maldistributed flow at low flow rates (top, shaded red) or equally distributed flow for all flow rates (bottom, shaded green). However, it is unclear what flow distribution may occur in parallel channels and why.

Previous studies have used a criterion based on the perturbation of the momentum balance equation to distinguish between stable (feasible) and unstable (infeasible) flow distributions. For example, based on this criterion, the stable distribution for the above system has been predicted as a maldistributed flow rather than a uniformly distributed one. While this prediction may be accurate, the criterion provides no reason why a maldistributed state is more stable than equally distributed flow when both distributions satisfy the basic equations governing mass, momentum, and energy conservation. Moreover, the number of viable flow distributions increases with the number of parallel channels, making the traditional modeling approach more challenging – hence, a clearer understanding of the maldistribution phenomenon is necessary.

The challenge introduced by non-unique solutions can be addressed by considering the entropic aspects of flow, which is often overlooked. Internal flow with energy transfer is associated with entropy generation as flows occur from the inlet $i$ to the exit $e$ of a channel. For steady flow through boundary-heated channels, the total entropy generation rate, $\dot{S}_{gen}$ can be calculated using the entropy balance equation for an open system.

$$\int_{CS} s\rho \bar{v} \cdot d\bar{A} = \int_{CS} \frac{d\dot{Q}}{T_w} + \dot{S}_{gen} \qquad (4)$$

Here $\dot{Q}$ denotes net heat transferred from the boundary into each channel maintained at a uniform temperature of $T_w$. The increase in entropy is also caused by the irreversibility associated with flow in the channels, $\dot{S}_{gen}$. Irreversibility occurs in this pressure-driven flow due to many reasons, including heat transfer across a finite temperature difference, friction between the fluid and the channel wall, sudden changes in configuration and direction of flow, and the mixing of thermally dissimilar fluids from parallel channels at the flow junctions or headers [20]. The second law of thermodynamics requires that the rate of entropy generation, $\dot{S}_{gen} \geq 0$, wherein equality holds only if the process is internally reversible. We hypothesize that the flow distribution in a multi-channel network is governed by the maximization of $\dot{S}_{gen}$. With the help of experiments involving a two parallel thermally-isolated channel flow system, we show how entropy analysis determines the thermodynamically preferred flow distribution, which matches our observations.

EXPERIMENTAL PROCEDURE

The test section (Figure 3a) comprises two parallel channels with shared inlet and exit. Each channel consists of a valve to control flow, a flowmeter (Omega FLR-1008ST, $\pm 3.3 \times 10^{-5}$ kg/s), and channels made of stainless steel capillary tubes of internal diameter $1.4 \times 10^{-3}$ m, outer diameter $3.18 \times 10^{-3}$ m and length $3.05 \times 10^{-1}$ m, circumferentially interfaced with heaters (125 W) and fiberglass insulation. Also, pressure sensors (Omega PX309-030A5V, $\pm 0.52$kPa) and thermocouples (Omega T-type, $\pm 1$°C) were positioned at the inlet and exit of the test section to monitor flow. Four additional temperature sensors were attached to each channel wall at equidistant positions to keep track of the channel wall temperature. Other components in the testbed include a temperature-controlled heated tank as a liquid reservoir, a chiller to extract heat from the working fluid, an electronic valve , and a gear pump (Micropump, GA-X21.CFS.E, 0 – 65 Hz) to maintain stable two-phase flow in the channels. The parallel channels are independently heated using programmable DC power supply modules (BK Precision, XLN 15010, $\pm 0.01$%). The testbed uses water, and the system was maintained within a pressure range of 8 to 33kPa.

In order to improve baseline stability,ensure consistency and maintain the required pressure at the start of an experiment, the setup is initially de-aerated, and the fluid tank is heated before turning on the pump. This step ensures that the working fluid instantaneously gets to the pump at startup, preventing cavitation and potential damage to the pump. Experiments typically involve liquid flow in sequence from the pump to the test section, storage tank, and chiller. Within the test section, fluid may split between two parallel channels, or flow might be limited to a single channel if the valve on the other branch is fully shut. Experiments typically involve liquid entering the test section below saturation conditions and exiting either as liquid, liquid vapor mixture, or super-heated vapor.

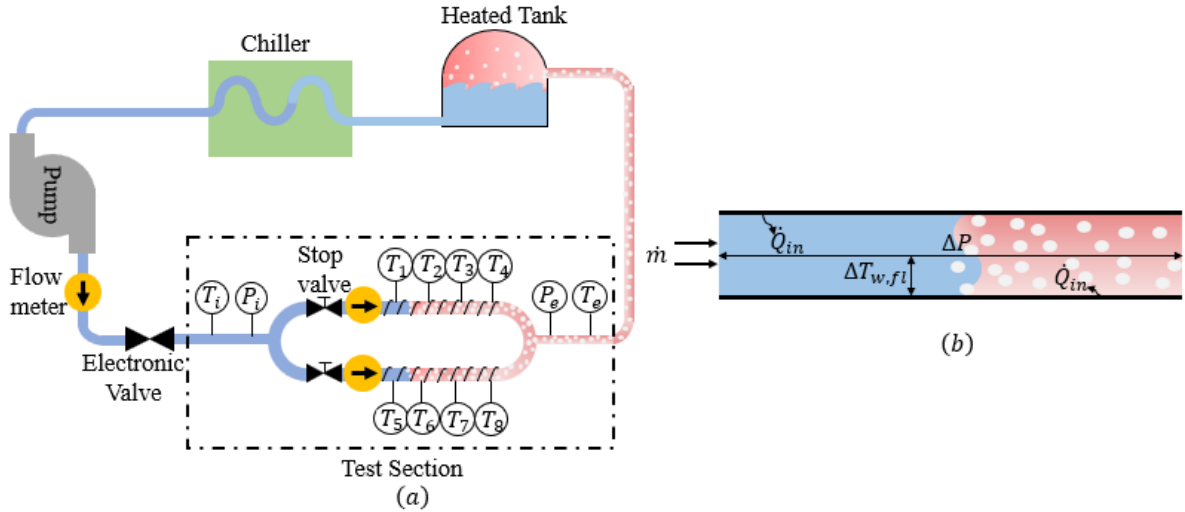

Figure 3. ($a$) Experimental pumped liquid cooling cycle. ($b$) Illustration of flow in a single channel.

Initial experiments involve thermal characterization of each channel to determine the relationship between readily available physical quantities (mass flow rate $\dot{m}$, channel surface area $A_s$ and temperature difference between the channel wall and ambient $T_w - T_\infty$) and heat loss $\dot{Q}_{loss}$ to the surroundings, as shown in Figure 4. Therefore, for a given electrical heat load $\dot{Q}_h = IV$, the net heat transfer to the fluid $\dot{Q} = \dot{Q}_h - \dot{Q}_{loss}$, where $\dot{Q}_{loss} = \dot{Q}_{loss}(\dot{m}, A_s, T_w - T_\infty)$ can be estimated. This step is useful, especially when $\dot{Q}$ cannot be directly obtained using measurements of mass, temperature, and pressure of the liquid-vapor mixture flow. An interface designed with the aid of LABVIEW and MATLAB is used for automated actuator control and data collation.

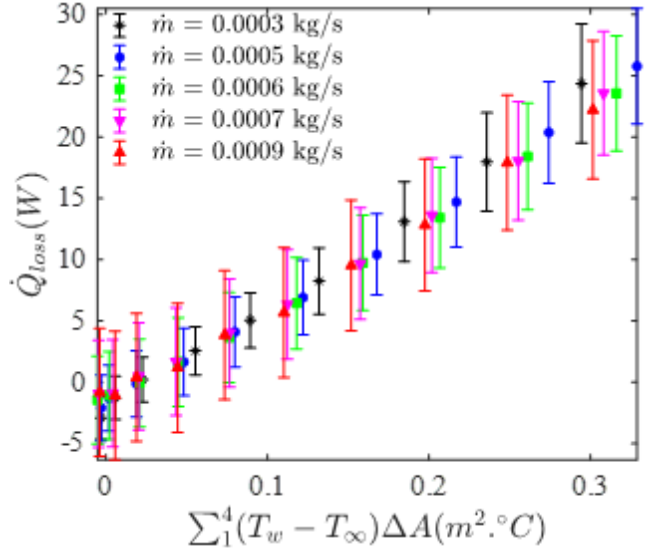

Figure 4. Heat loss characterization of a single channel

Uncertainty $\Delta f$ quantification, is carried out by applying the propagation of error [21]–[23] described in Eq. 5 on experimental data.

$$\Delta f = \sqrt{\left(\frac{\partial f}{\partial x_1}\Delta x_1\right)^2 + \left(\frac{\partial f}{\partial x_2}\Delta x_2\right)^2 + \ldots + \left(\frac{\partial f}{\partial x_n}\Delta x_n\right)^2} \qquad (5)$$

Where $\Delta x_1 \ldots \Delta x_n$ are the errors associated with the measurement of $n$ independent variables $x_1 \ldots x_n$ of the function $f$.

ENTROPY GENERATION AND SOURCES OF IRREVERSIBILITY

The following equations describe the first and second laws of thermodynamics for the N-channel system, which allow for determining $\dot{S}_{gen}$.

$$\dot{m}(h_i - h_e) = \sum_{j=1}^{N} \dot{Q}_{h,j} - \dot{Q}_{loss,j} \qquad (6)$$

$$\dot{S}_{gen} = \dot{m}\big(s_e(P_e, h_e) - s_i(P_i, T_i)\big) - \sum_{j=1}^{N} \frac{\dot{Q}_{h,j} - \dot{Q}_{loss,j}}{T_{w,j}} \qquad (7)$$

The specific entropies at the inlet ($s_i$) and outlet ($s_e$) of the channel assembly are functions of inlet pressure $P_i$ and temperature $T_i$, and exit pressure $P_e$ and specific enthalpy, $h_e$, respectively. Here $\dot{Q}$ is obtained from the difference between the electrical heat load $\dot{Q}_{h,j}$ and heat loss $\dot{Q}_{\infty,j}$ in each channel $j$. In a system of parallel channels, the total rate of entropy generated consists of the rate of entropy generated within each channel member $\dot{S}_{gen,j}$, the rate of entropy generated from the splitting of flow at the common inlet $\dot{S}_{gen,split}$ and the rate of entropy generated at the common exit from irreversible heat transfer within mixing fluid streams $\dot{S}_{gen,mix}$. Compared with other contributions, $\dot{S}_{gen,split}$ is negligible, because of the insignificant change in the thermodynamic properties of the fluid during the splitting process. As a result, $\dot{S}_{gen}$ is expressed as

$$\dot{S}_{gen} = \dot{S}_{gen,mix} + \sum_{j=1}^{N} \dot{S}_{gen,j} \qquad (8)$$

Since the channel assembly is thermally insulated and the mixing process is relatively fast, adiabatic conditions are assumed in determining $\dot{S}_{gen,mix}$.

$$\dot{S}_{gen,mix} = \dot{m} s_{mix}(P_{mix}, h_{mix}) - \sum_{j=1}^{N} \dot{m}_j s_{e,j} \qquad (9)$$

Where the specific entropy of the fluid mixture at the exit ($s_{mix}$) is a function of the mixture pressure $P_{mix}$ and enthalpy $h_{mix}$. In parallel channel experiments, $P_{mix}$ and $h_{mix}$ are typically $P_e$ and $h_e$ respectively. However, in cases where single channel experiments are used to estimate the properties of flow in parallel channels, $P_{mix}$ and $h_{mix}$ are given as

$$h_{mix} = \sum_{j=1}^{N} \dot{m}_j^* h_{e,j}, \tag{10}$$

$$P_{mix} = \sum_{j=1}^{N} \dot{m}_j^* P_{e,j}. \tag{11}$$

## A. Sources of Irreversibility in a Single Channel

Besides other sources, the hydraulic and th`ermal resistance affecting the flow and heat transfer in the two channels contribute significantly to irreversibility. The pressure drop $\Delta P = P_i - P_e$ across the channel depends on the flow rate and the hydraulic resistance to flow, which depends on fluid friction. Hence, $\Delta P$ indicates how certain flow aspects (e.g., friction) introduce irreversibility in the system. Similarly, heat transfer from the wall into the fluid requires a temperature differential. For a hypothetical reversible heat transfer process ($\dot{S}_{gen} = 0$), the boundary or channel wall temperature ($T_{rev}$) can be calculated using the following equation.

$$T_{rev} = \frac{\dot{Q}_h - \dot{Q}_{loss}}{\dot{m}(s(P_i, h_e) - s(P_i, T_i))} \tag{12}$$

However, the actual channel wall temperature ($T_w$) is much larger than $T_{rev}$. Hence, $\Delta T_w = T_w - T_{rev}$ is indicative of the deviation from an ideal heat transfer process. Figure 5a shows $\Delta P$, $\Delta T_w$, the exit vapor quality $x_e$, and the entropy generation rate $\dot{S}_{gen}$ for a single channel heated at $\dot{Q}_h = 70$ W.

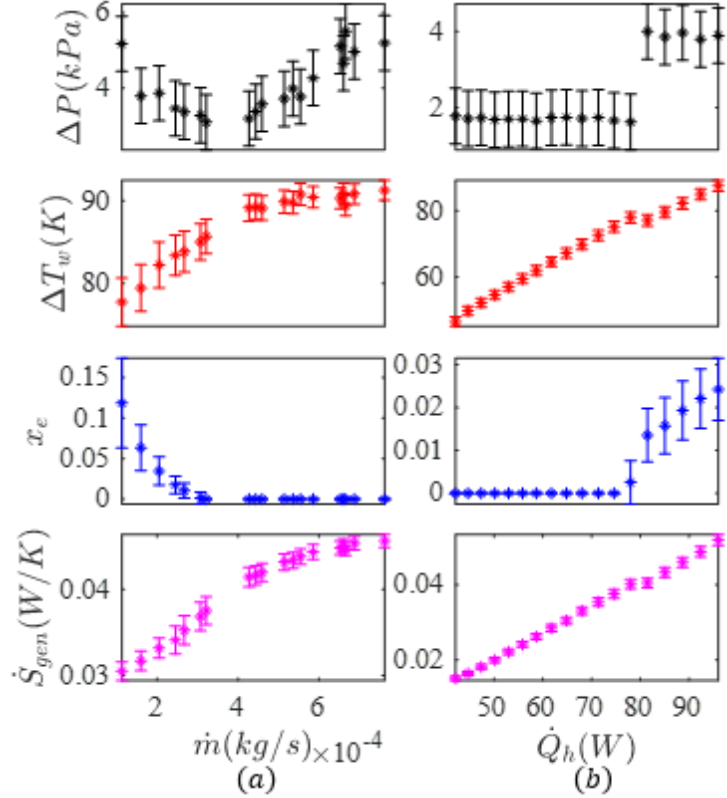

Figure 5. Variation of single-channel pressure drop $\Delta P$, wall temperature difference $\Delta T_w$, exit quality $x_e$, and entropy generation $\dot{S}_{gen}$, as a function of (a) flow rate $\dot{m}$ at $\dot{Q}_h$= 70 W. (b) external heat load $\dot{Q}_h$ at $\dot{m} = $ 3.48 kg/s.

Starting from a large flow rate, a continuous decrease in $\dot{m}$ results in liquid-to-vapor phase change. Consequently, the fluid phase at the exit changes from liquid ($x_e = 0$) to a liquid-vapor mixture ($x_e > 0$). With phase-change and increase in vapor production, heat transfer characteristics within the channel change to become more efficient, which can be seen in the lowering of $\Delta T_w$ with an increasing $x_e$. Correspondingly, the rate of entropy generation also decreases. This decrease in $\dot{S}_{gen}$ occurs despite increasing $\Delta P$, which indicates the more significant role of heat transfer irreversibility over flow non-idealities. At large flow rates, when $\Delta T_w$ does not change much with an increase in $\dot{m}$, $\dot{S}_{gen}$ increases mainly due to rising $\Delta P$ – a regime where flow non-idealities are dominant.

To further investigate the contribution of flow and heat transfer irreversibilities to the total entropy production, we also varied channel heating rates ($\dot{Q}_h$) while keeping the flow rate ($\dot{m}$) constant. Figure 5b shows the variation of $\Delta P$, $\Delta T_w$, the exit vapor quality ($x_e$), and the entropy generation rate $\dot{S}_{gen}$ for a single channel with $\dot{Q}_h$ at a constant $\dot{m} = 3.48$ kg/s. Starting from a small heat load, a continuous increase in $\dot{Q}_h$ results in the phase change from liquid ($x_e = 0$) to liquid-vapor mixture ($x_e > 0$), a jump in $\Delta P$ and an increase in both $\Delta T_w$ and $\dot{S}_{gen}$. The sudden increase in $\Delta P$ is due to vapor production (phase change) in the channel at the threshold value of the heating rate around 80 W. The increase in $\Delta T_w$ and $\dot{S}_{gen}$ with no significant change in $\Delta P$ before and after phase change indicates that the contribution to $\dot{S}_{gen}$ of the irreversibilities associated with the heat transfer process are more significant than those associated with the flow.

## B. Irreversibility in two thermally isolated Parallel Channels

Unlike flow in a single channel, entropy generation rates in two parallel channels are also influenced by the nature of the flow distribution. Flow distribution is quantified in terms of $\dot{m}^*$ which denotes the flow fraction in a channel relative to the total flow rate. Flow distribution is obtained based on measured flow rates, $\dot{m}_1^* = \dot{m}_1/(\dot{m}_1 + \dot{m}_2)$ and $\dot{m}_2^* = \dot{m}_2/(\dot{m}_1 + \dot{m}_2)$. For instance, $\dot{m}^* = 0.5$ denotes equally distributed flow in a two-channel system. Figure shows the observed flow fraction $\dot{m}^*$, total entropy generation $\dot{S}_{gen}$, pressure drop $\Delta P$ and the temperature difference $\Delta T_w$ for different total flow rates $\dot{m}$ in two nominally identical thermally isolated parallel channels under a constant heat load of 70 W each.

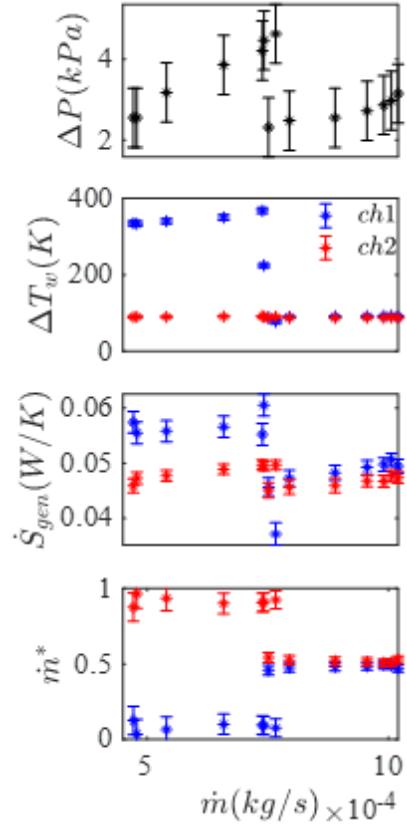

Figure 6. Variation of parallel channel pressure drop $\Delta P$, wall temperature difference $\Delta T_w$, entropy generation $\dot{S}_{gen}$ and flow fraction $\dot{m}^*$ as a function of the flow rate $\dot{m}$ at $\dot{Q}_h= 70$ W

At high flow rates in the two parallel identical channels, the pressure difference ($\Delta P$) decreases as the total flow rate ($\dot{m}$) decreases. For a large $\dot{m}$, the fluid emerges as a liquid at the exit. Since the characteristics of heat transfer do not change much under these conditions, there is no significant change in the temperature difference, $\Delta T_w$. Hence, like flow in a single channel, the $\dot{S}_{gen}$ in each channel, and the $\Delta P$ across the channel decreases as $\dot{m}$ decreases and $\Delta T_w$ is expectedly uniform.

As $\dot{m}$ decreases further, fluid at the exit transitions from liquid to a liquid-vapor mixture due to phase-change caused by heating. Phase-change characteristics now initiate severe flow maldistribution, as indicated by the departure in $\dot{m}^*$ for the channels. This uneven distribution occurs with a sudden decrease in $\dot{m}^*$ for channel 1 and increase in $\dot{m}^*$ for channel 2. A sudden rise in $\Delta P$ across the channels is also observed. The decrease in $\dot{m}$ also results in a sharp rise in $\Delta T_w$ and $\dot{S}_{gen}$ in channel 1. In comparison, the change in $\dot{S}_{gen}$ is not that significant and $\Delta T_w$ is mostly invariant in channel 2.

The relative effects of flow and heat transfer irreversibilities on $\dot{S}_{gen}$ depend on the system. In this case, when the flow was equally distributed, or when it remained maldistributed, the variation in $\dot{S}_{gen}$ is due to hydraulic irreversibilities. However, in this case, the transition from uniform to non-uniform flow distribution was mainly due to thermal irreversibilities. Considering both channels collectively, there is an increase in the total $\dot{S}_{gen}$. Hence, an important conclusion from this observation is that severe flow maldistribution is associated with a sharp rise in entropy production. The following section describes why severe maldistributed flows prevail over uniform and moderately non-uniform flows.

## Entropy Generation and Flow Distribution in thermally isolated Parallel Channels

Although the flow rate in a single channel can be predicted based on the pressure drop and the channel geometry [1], the possible flow distributions in parallel channels are multiple and not easily predictable. This section shows why severely maldistributed flow prevails over uniform and moderately non-uniform distribution by considering the two parallel channels. Here, we compare the entropy generation in an actual flow distribution with the entropy generation if the flow were to be uniformly distributed.

### A. *Flow in Uniform Parallel Channels*

The parallel channel assembly shown in Figure 3 is used to compare the entropy generation rates for severely maldistributed and uniformly distributed flows. In order to calculate the entropy production rates for maldistributed flow, the two channels are equally heated ($\dot{Q}_{h,1} = \dot{Q}_{h,2} = 70$ W) and the total flow rate, $\dot{m}$ is gradually reduced (Figure 7a). Alternately, the total flow rate can be held fixed ($\dot{m} = 6.96 \times 10^{-4}$ kg/s) and the heating rates, although uniform ($\dot{Q}_{h,1} = \dot{Q}_{h,2} = 0.5\dot{Q}_h$), can be gradually increased (Figure 7b). Using the measured inlet and exit flow properties and wall temperatures, Eq. (7) is used to determine $\dot{S}_{gen}$. The entropy generation rate and flow distributions are indicated as "coupled" channels in Figure 7.

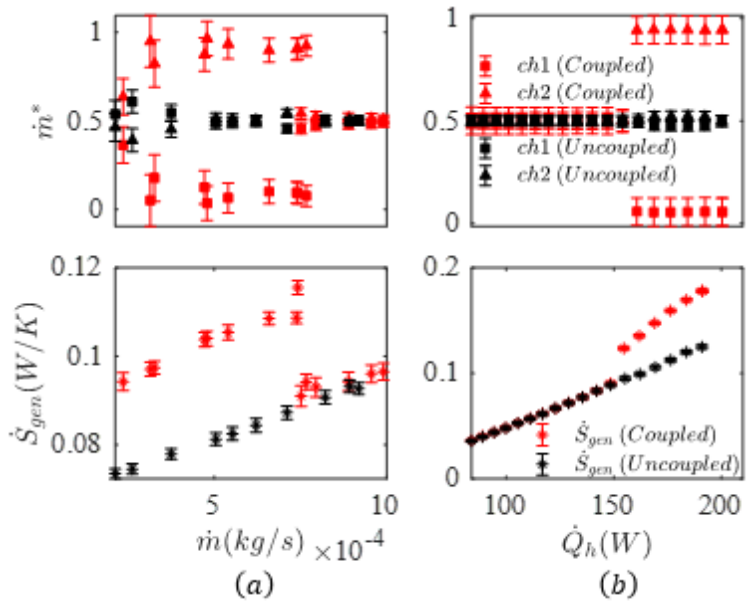

Figure 7. Variation of $\dot{m}^*$ and $\dot{S}_{gen}$ with $\dot{m}$ for severely maldistributed and uniformly distributed flow for (a) varying $\dot{m}$ with $\dot{Q}_{h1} = \dot{Q}_{h2} = 70\ W$, and (b) varying $\dot{Q}_h$ at $\dot{m} = 6.96 \times 10^{-4}$ kg/s.

Estimating the entropy generation rate for uniformly distributed flow in parallel channels is not straightforward because uniform flow does not occur naturally under these conditions. For estimation, we individually test each channel in the testbed by isolating the other channel from the system using stop valves shown in Figure 3. Due to similar geometry and heating conditions, the total rate of entropy generation is obtained from Eq. 8 based on a common experimental input variable such as flow rate ($\dot{m}_1 = \dot{m}_2 = \dot{m}/2$) (Figure 7a) or heat load ($\dot{Q}_{h,1} = \dot{Q}_{h,2}$) (Figure 7b), yielding $\dot{S}_{gen}$ corresponding to the hypothetical case of uniformly distributed flow ($m^* = 0.5$) in the parallel channel assembly. These results are shown as "uncoupled" channels in Figure 7.

Figure 7 compares $\dot{S}_{gen}$ for maldistributed and uniformly distributed flow in the parallel channels. Figure 7a shows flow distribution and entropy generation when the flow rate $\dot{m}$ is gradually decreased, while a steady heating power of 70 W is applied to both channels. Likewise, Figure 7b shows the flow distribution and entropy production when the heating load, although the same for both channels, is gradually increased with a total flow rate fixed at $\dot{m} = 6.96 \times 10^{-4}$ kg/s. In case of single channel experiments, the flow rate in each channel is halved with $\dot{m} = 3.48 \times 10^{-4}$ kg/s.

When $\dot{m}$ is large or $\dot{Q}_h$ is low, the flow is equally distributed between the channels. For these conditions, $\dot{S}_{gen}$ decreases with $\dot{m}$ (Figure 7a) and increases with $\dot{Q}_h$ (Figure 7b). However, as $\dot{m}$ decreases or $\dot{Q}_h$ increases, the flow eventually becomes severely maldistributed. As evident in Figure 7, this maldistribution corresponds to a sudden jump in $\dot{S}_{gen}$. However, if the channels experience uniform flow, then $\dot{S}_{gen}$ does not show any dramatic variation. For these conditions representing a hypothetical uniform flow case, $\dot{S}_{gen}$ continues to decrease with $\dot{m}$ and increase with $\dot{Q}_h$. Figure 7 shows that entropy generation rate of severely maldistributed flow is much greater than equally distributed flow under similar conditions. Hence, flow maldistribution seen here is thermodynamically more favorable over equally distributed flow. It is due to this reason that non-uniform flows are sustained unless conditions are significantly altered to enable uniform flow distribution.

## B. Flow in Non-uniform Parallel Channels

Severe maldistribution also occurs in non-uniform or dissimilar channels, which can be explained using entropy analysis. To understand the effects of channel dissimilarity, we use different heat loads and valve openings associated with each channel to compare the entropy production rates. Figure 8 shows the flow behavior with $\dot{Q}_{h,1} = 80$ W and $\dot{Q}_{h,2} = 70$ W, with channel 1 valve fully open and the channel 2 valve partially closed. Since the channels are inherently different, flow distribution is never uniform. Still, model predictions could indicate multiple possibilities. For example, in this case, the predictions can indicate either marginal or significant maldistributions in these channels.

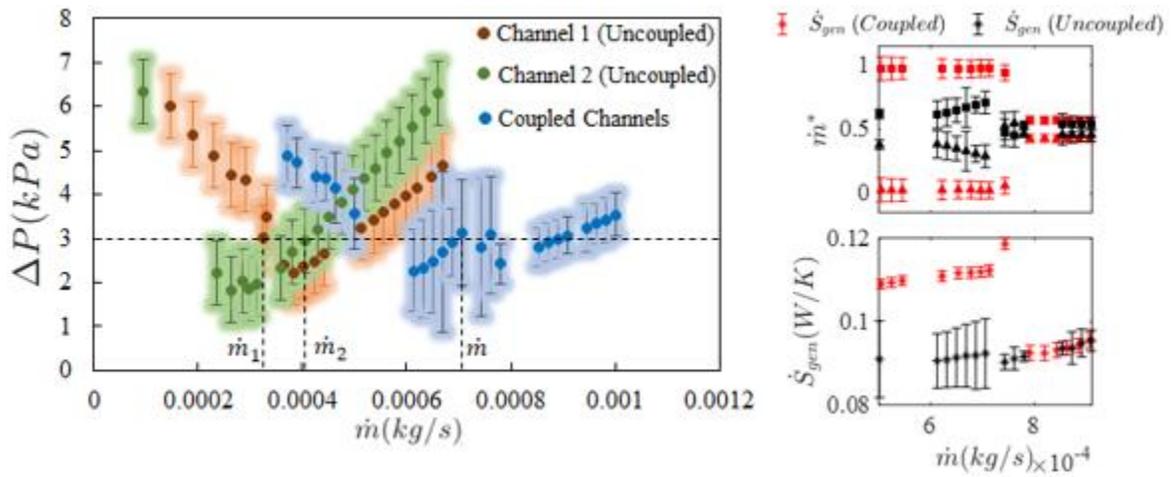

Figure 8. Variation of $\dot{m}^*$ and $\dot{S}_{gen}$ with $\dot{m}$ for severely maldistributed and moderately maldistributed flow

When flow occurs in both channels simultaneously, the distribution ($\dot{m}^*$) is directly measured, and Eq. (7) is used to obtain $\dot{S}_{gen}$ using the measured flow properties as described earlier. These results shown in Figure 8 (right) are indicated as coupled channels. In order to estimate other possible distributions, we use flow characteristics in individual (uncoupled) channels, with heating and valve settings similar to the coupled channels experiment. The flow characteristics, given by $\Delta P$ versus $\dot{m}$ for the uncoupled and coupled channels, are shown in Figure 8 (left). For a total flow rate $\dot{m}$ observed in the experiments, the flow distributions in uncoupled channels 1 ($\dot{m}_1$) and 2 ($\dot{m}_2$) are obtained such that $\dot{m}_1 + \dot{m}_2 = \dot{m}$, with flow rates $\dot{m}_1$ and $\dot{m}_2$ corresponding to the same pressure drop $\Delta P$ for the specific $\dot{m}$ in the coupled experiments. To do this we find the intersection of a constant $\Delta P$ line with the $\Delta P$ versus $\dot{m}$ characteristic curves for each channel, resulting in flow distribution $\dot{m}_1^*$ and $\dot{m}_2^*$ whose sum is unity. Mathematically, this can be expressed as a minimization problem with $\dot{m}_1^*$ as the argument.

$$\dot{m}_1^* = arg \min_{\dot{m}_1^* \in [0,1]} |\Delta P_1(\dot{m}_1^*) - \Delta P_2(1 - \dot{m}_1^*)| \tag{13}$$

This procedure allows calculating $\dot{m}_1^*, \dot{m}_2^* = 1 - \dot{m}_1^*$ and the entropy generation, $\dot{S}_{gen}$ using $\dot{S}_{gen,1}$ and $\dot{S}_{gen,2}$ from single channel experiments, and $\dot{S}_{gen,mix}$ to account for the adiabatic mixing of the two fluid streams in Eq. (8). In conclusion, the resulting flow distribution for uncoupled channels indicate a moderate maldistribution, and the corresponding $\dot{S}_{gen}$ is less than the coupled experiment.

Due to dissimilar channel geometries and heating, flow characteristics are expected to differ. For large values of $\dot{m}$ both uncoupled and coupled channel results indicate that the flow is slightly maldistributed mainly due to differences in channel geometry and $\dot{S}_{gen}$ decreases with $\dot{m}$. But as the total flow rate $\dot{m}$ decreases further, both coupled and uncoupled channel results indicate larger flow maldistribution. Typical static modeling approaches could indicate that moderate maldistribution distribution is viable even for low total flow rates. However, an entropy analysis shows that the $\dot{S}_{gen}$ corresponding to severely maldistributed flow is greater than the $\dot{S}_{gen}$ corresponding to the moderately maldistributed flow. Consequently, the significantly maldistributed state is thermodynamically favored over moderately non-uniform flow, as seen in the experiments.

CONCLUSION

This study analyzes entropy production and its relationship with flow distribution in thermally isolated parallel channels with multiphase flow. Parallel channels with multiphase flow, even if nominally identical, can experience maldistribution or unequal flow distribution, which is undesirable in many applications. Since such systems are inherently non-linear due to multiphase flow characteristics, modeling and predicting this phenomenon is challenging and often results in multiple viable solutions. This study shows how an entropy analysis of flow in channels can overcome this modeling challenge. We hypothesize that flow maldistribution is governed by a thermodynamically favored state with maximum possible entropy production among the several flow possibilities identified by solving the fundamental equations governing mass, momentum, and energy conservation.

Using experiments, we identify two primary sources of irreversibilities associated with heat transfer and flow. Although both sources are present under all conditions, experiments show that flow irreversibilities, such as flow friction, affect entropy generation at higher flow rates. In contrast, thermal irreversibilities, such as non-isothermal heat transfer, tend to affect entropy generation more at lower flow rates and higher heat rates.

The flow distribution in parallel nominally identical channels is uniform for high flow rates, corresponding to little to no phase change. However, a transition to maldistribution occurs when the flow rate is reduced, or the heating rate is increased. The transition from uniform to non-uniform flow distribution corresponds to a sharp rise in entropy production. Such an increase is not predicted for uniform flow distribution across parallel channels, making maldistributed flow a thermodynamically favored state over equally distributed flow. Maldistributed flow corresponds to the highest entropy production rate compared to other steady two-phase flow distributions for similar system constraints. Such flow behavior also exists in non-identical channels, wherein moderately non-uniform flow distributions could be predicted. However, in practice, the flow distribution is more significantly non-uniform since it is thermodynamically favored and corresponds to a higher entropy production rate.

Entropy analysis is often omitted from the modeling and performance prediction of systems with multiphase flow and liquid-vapor phase change. This study shows how entropic considerations provide an additional perspective to understand and predict the behavior of non-linear systems with several possible outcomes. Although this study analyzes stable and unstable flow distribution in two thermally isolated parallel channels, the principle of maximum entropy production may be extended to analyzing a larger system of parallel channels experiencing heating and multiphase flow or even thermally coupled channels.

ACKNOWLEDGMENTS

The work was supported by the National Science Foundation's Division of Chemical, Bioengineering, Environmental, and Transport Systems in the Directorate for Engineering under Grant No. 1944323.